\documentclass[aps,prd,preprint,groupedaddress,showpacs]{revtex4-1} 
\usepackage{graphicx}
\begin{document}
\title{Time Symmetric Quantum Mechanics\\ and
Causal Classical Physics\,?}
\author{Fritz W. Bopp}
\affiliation{University of Siegen}

\date{\today}

\begin{abstract}
A two boundary quantum mechanics without time ordered causal structure
is advocated as consistent theory. The apparent causal structure of
usual ,,near future`` macroscopic phenomena is attributed to a cosmological
asymmetry and to rules governing the transition between microscopic
to macroscopic observations. Our interest is a heuristic understanding
of the resulting macroscopic physics.
\end{abstract}

\pacs{03.65.Ta, 03.65.Ud}

\maketitle

\section*{Introduction}

There are a number of paradoxes attributed to quantum mechanics involving
the transition to macroscopic physics. As pointed out by Einstein,
Podolsky, and Rosen\,\cite{einstein1935can} a ,,collapse`` of
the wave function seems to violate the local structure of the theory.
There are a number of other odd features connected to the measurement
process\,\cite{schrodinger1926stetige,schrodinger1935gegenwartige,wigner1961remarks}.
They are widely discussed in the extensive literature on the philosophy
of quantum mechanics (see e.g.\,\cite{bohr1935can,bell1964einstein,mermin1998quantum,barrett1999quantum,friebe2014philosophie}). 

These discussions neglect an in our opinion more illuminating paradox,
which relies on a careful consideration of the Hanbury-Brown Twiss
interferometry\,\cite{brown1957interferometry}. The Hanbury-Brown
Twiss interferometry - or the time when it was generally accepted
- is comparatively young. It also involves quantum statistics.

To resolve the paradox a quantum mechanical world without time ordered
causal structure with a fixed initial and fixed or strongly restricted
final state is conjectured. Our basic idea is that there is no problem
in the backward causation, if there is a \textit{{way
to}} \textit{{restore causality }}{in
the transition to the usual known part of the  macroscopic world}. 

The aim of this note is to better understand how such a restoration
could work. On general terms two obviously needed rules for the transition
to macroscopia\textit{{\emph{ are formulated}}}.
In section~3 and 4 we turn to a general discussion of a quantum mechanics
with two boundary state vectors following the work of Aharonov and
coauthors\,\cite{aharonov1964time,aharonov2008quantum}. 
A cosmological
consideration similar to an idea of Gell-Mann and
Hartle\, \cite{gell1994time} follows in section~5 and 7. The paper argues how the cosmological
expansion allows for something which locally looks like a collapse  structure and effectively
introduces a time arrow. How in such a scenario
a time-ordered causal macroscopia could arise, is outlined with a
simple router picture and a small toy program in the central section~6.

\section*{1\hspace*{0.5cm}Argument for backward causation}

Following Hanbury-Brown Twiss (figure~1) we consider a star emitting
two photons with equal frequency, polarization and phase in direction
of an observatory with two telescope detectors. The star should be
light years away and an attribution of the photons to the closely
neighboring detectors $(\mathrm{separation}\,\Delta\to0)$ should
not be possible. If the observatory happens to observe them the interference
term leads to a quantum statistical enhancement of the emission probability
by a factor two. If it mirrors them back in space - with a mirror
large enough to  allow for a resolution of the positions of the
emitters - no such factor occurs. So the choice affects a probability
of an event way past.
\begin{figure}
\noindent \centering{}\includegraphics[scale=0.6]{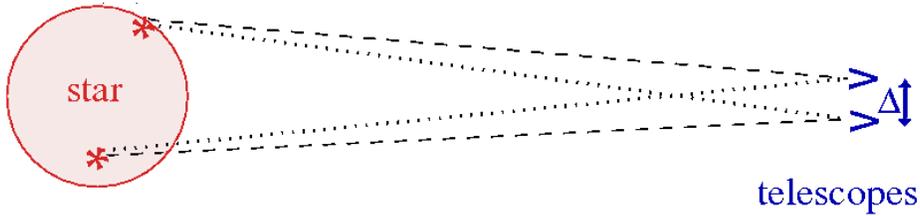}\protect\caption{Hanbury-Brown Twiss observation}
\end{figure}

Of course there is the opposite phase case where a corresponding equal
quantum statistical suppression occurs. In this way summing up both
situations the observation choice does not affect the total emission
probability. It just affects the wave function in the past by enhancing
its same phase component. Such backward causation in wave functions
is known from Wheelers delayed-choice gedanken 
experiment\,\cite{wheeler1984quantum,wheeler2014quantum,hellmuth1987delayed,ma2014delayed}.

However the emission process here doesn't have to be incoherent and
phases can be correlated in the emission process. We therefore conclude
that there can be special situations where past emission probabilities
are changed. The causal direction is broken independent of the ontological
role of wave functions\,\cite{bopp2001ulm}. 

We stress that the observation of the second order Hanbury-Brown Twiss
correlation and devices with coherent emission are real and not gedanken
constructs. The conclusion is not optional. What follows is not another
interpretation of quantum mechanics but anattempt to encounter this
observation.

Similar observations exist in other contexts. Consider multi-particle production 
in nuclei or particle scattering. A quantum statistical enhancement 
in the production of very closely neighboring identical bosons and a
corresponding suppression in
that of very closely neighboring identical fermions was observed
as peak or dip against a smoother background\,\cite{kittel2005soft}.
The enhancement or suppression of the emission obviously could be eliminated 
after the initial scattering if a third particle crosses the paths of the
pair. The crossing had to be in a region where the different  emission
regions of both particles would still  be distinguished.  

$ $   

In atomic physics it could be shown that the absorption probability of
a photon by an atom can be drastically increased by a parabolic
mirror\, \cite{quabis2000focusing,sondermann2007design}. In the reverse process 
the emission probability depends on the presence of the parabolic
mirror which can be far away and thus manipulated at a later time. 
Again there is a backward causation effect on a 
probability (see also\,\cite{drexhage1974progress,walther2006cavity}). 
I am aware that the absence of a microscopic causal structure  is not 
widely accepted and this is a point where more real experiments
would be persuasive. 

The backward causation of emissions contradicts the De-Broglie-Bohm
theory\,\cite{broglie1927structure,bohm1957discussion,durr2009bohmian}.
If one photon is observed in the first telescope Bohm’s law of motion
($\psi$ is the usual wave function.)
\[
\left.\left(\psi^{*}\psi\right)\cdot d\overrightarrow{Q}/dt=\frac{\hbar}{m_{i}}\left(\psi^{*}\nabla\psi\right)\right|_{\mathrm{for\, all\, points\, on\, path}\, Q(t)}
\]
can non-locally guide the path $Q(t)$ of the second particle to the
second telescope with the required enhanced probability for the second
order correlation. But in the manifest forward evolution past emission
probabilities cannot be affected. 

Can the broken causal direction be understood in usual quantum mechanics?
Quantum mechanics knows the amplitude that a given initial state evolves
to a given final one. The calculation uses a collapse-less theory (Sakurai
uses the name ,,quantum dynamics`` \cite{sakurai2014modern}) which
contains no intrinsic time direction. No contradiction to quantum
dynamics arises.

The absence of a time direction in basic laws is intuitively annoying and
many authors tried to introduce dynamical time arrows. In field theory it is
possible to basically admit only advanced solutions and create in this way an
asymmetry
\cite{ritz1908grundlagen,tetrode1922wirkungszusammenhang,frenkel1925elektrodynamik}. 
However this apparent asymmetry gets lost
in the path integral formalism of Feynman \cite{feynman1948space}. The hypothesis adhered
to here is that the dynamical equations are time invariant and that there is no
quantum dynamical time arrow. It sides with Einstein in its controversy with
Ritz \cite{ritz1909gegenwartigen}  and follows many outstanding authors
like~\cite{wheeler1984quantum}. 
For a detailed consideration of the time arrows we refer to Zeh's
book \cite{zeh2001physical}.  

\vspace{3mm}
\section*{2\hspace*{0.5cm}\textit{{\emph{Correspondence transition
rules}}}}

The backward causation has to be restricted to microscopic phenomena.
A key point is what has to be taken as microscopic and macroscopic.
The example of backward causation discussed in the beginning  tells us that the
initial state in the ,,macroscopic`` (non quantum) world cannot
contain correlated phases. In this way the example with locked-in
phases has - albeit the possible extension of the process - to be
attributed to the ,,microscopic`` (quantum) world.

Backward causation in the microscopic world involves interference
effects. The absence of backward causation in the macroscopic world
indicates:
\begin{itemize}
\item Macroscopically prepared initial states contain no correlated phases. 
\item Macroscopic measurements obtain an equal contribution from states
with quantum statistical enhancement and suppression. 
\end{itemize}
We denote this as \emph{,,correspondence transition rule}s`` (i.e.
rule of the transition to macroscopia described by the correspondence
principle). The cause of the rules is macroscopically unavoidable
phase averaging in the initial state and a macroscopically unavoidable
averaging over enhancing and suppressing contribution in final state
measurements. 

The second rule claims it is not possible to obtain enhancement or
suppression through subsequent interference effects. It is called
„no post-selection“\,\cite{zeh2001physical} by macroscopic devices.
To observe interferences one has to join distinct paths. 
As a consequence of the quantum Liouville theorem any device cannot 
reduce the number of available paths. The joining essentially
works like a partially reflecting mirror. Ignoring irrelevant aspects there are two incoming paths and
two outgoing ones. For given photon-phases $\phi_{1}$ and $\phi_{2}$
there can be a suppression like $\sin^{2}(\phi_{1}-\phi_{2})$ in
one of the directions and a corresponding enhancement like $\cos^{2}(\phi_{1}-\phi_{2})$
in the other. Macroscopically both channels have to be included and
the overall probability is not affected. 

The rules were developed for \textit{models of multi-particle production}
where peaks or dips against a smoother background are observed.
The rule postulates
that in an average over the neighboring peak or dip region and the not so neighboring region
enhancement and suppression effects occur with an equal weight (see\,\cite{andersson1986bose,metzger2011bose}
for an effective implementation in simulations codes). In this way 
quantum statistical enhancement or suppression doesn't destroy
the well tested factorization between the initial  and the hadronization
process. 
(The percent level determination of the QCD coupling constant from the 
hadronic structure in $e^+ e^-$- annihilation depends on this
factorization. No quantum statistical effect enhancing 
large multiplicity production is allowed.) 

The \textit{{\emph{correspondence transition rule}}}
plays an important role in the \textit{{understanding
of lasers}}. The lasing equation\,\cite{haken1985laser} is: 
\[
\frac{dn}{dt}=\left[(N_{2}-N_{1})\cdot W\cdot n\right]+\left[W\cdot N_{2}\right]-\left[2\cdot\kappa\cdot n\right]
\]where $N_{2}\,\mathrm{and}\, N_{1}$ are the occupation numbers of
the excited and the ground state, where $n$ is the number of photons,
where $W$ is the transition probability, and where $\kappa$ denotes
the absorption coefficient. It contains three terms (enclosed by rectangular
brackets): the stimulated emission or absorption, the spontaneous
emission, and the spontaneous absorption. To obtain the required exponential
growth in the photon number (lasing condition) a positive right side
is needed. 

This argument misses the crucial mechanism working in lasers. The
emission of a photon has - depending on the angular momentum of the
states - only a slight preference of the forward direction. The coherent,
precisely forward emission in the laser is a quantum statistical enhancement
not considered in above equation.

Why the lasing condition argument still makes sense is a consequence
of the\emph{ correspondence transition rule. }The lasing condition
is a purely macroscopic consideration just counting the number of
photons for which the quantum statistical enhancement and suppression
effects cancel and - in spite of their otherwise pivotal role - 
can safely be ignored.\,

\section*{3\hspace*{0.5cm}Two boundary quantum system}
\vspace{-3mm}

The argument above shows that the present can be affected by the past
and by the future. It means that two boundaries are needed
to describe the present situation and to avoid  an artificial time asymmetry.
A two boundary  quantum mechanics is  possible but it will significantly change the
picture\,\cite{schulman1997Time,price2012does,aharonov1964time,aharonov2010time,miller1996realism,reznik1995time,gell1994time}. 
Except for the matching procedure discussed below it is unitary.
It provides a completely consistent theory without paradoxes. 
No other change to the usual quantum dynamics is needed.
The apparent macroscopic time asymmetry will be attributed to our
cosmological situation.

Such a symmetric theory can avoid collapsing wave functions. 
Obviously quantum mechanics with a two directional constraint does
not contain locality problems like the EPR paradox\,\cite{goldstein2003opposite}.
Simply, if a state can change backward in time it can obviously also
change faster than light in mixed forward backward processes. 

't Hooft\,\cite{hooft1990quantization} recently advocated that a
suitable deterministic local cellular automaton theory could underlay
quantum mechanics. Any such underlying deterministic theory involves
a fixed predetermined final state and therefore shares our question
how a fixed final quantum state can coexist with a causal macroscopic
world. 

The strategy is to take these initial and final states far away so
that there is a region in between which is in principle known by quantum
dynamics and which is huge enough so that something like classical
aspects of the intermediate evolution appears within the closed system.
Without contact to an ,,outside`` there are no collapses. As in
Everett's interpretation\,\cite{everett1957relative} different multi-worlds
can contribute in between. But the fixed final state severely limits
the proliferation of such intermediate ,,universes``. The hypothesis
is that in the present situation coexisting paths essentially only
survive as usual ,,quantum effects`` on a microscopic level. 

The picture with the fixed final state will obviously in the end also
involve a modification\textbf{ }{of classical physics.
}As quantum dynamics is tested to a considerably higher precision
than classical mechanics, it should be taken as better known\,\cite{bopp2013werner}
and untested parts of {classical physics might be
modified.}

\vspace{3mm}
\section*{4\hspace*{0.5cm}Measuring processes within a closed system}

Two boundary quantum systems were investigated with great 
care\,\cite{gell1994time,aharonov2010time,aharonov2014foundations}.
We recap parts needed to understand the resulting macroscopic description. 

An essential element of a measurement in open systems is to ensure
that states observed with different eigenvalues can no longer interfere.
It prohibits reconstruction of the premeasured state and introduces
a time arrow. 

In a closed system the situation is actually quite similar. As in
the open system the evolving wave function contains manifest deterministic
parts with all kinds of interactions, including branching and merging.
The system also contains seemingly non deterministic measuring processes
corresponding to the usual measurements in open systems. A measured
subsystem is brought in contact with a witnessing subsystem\,\cite{zurek2003decoherence}
by a suitable interaction 
\[
H_{\mathrm{interaction}}=-g(t)\,\dot{p}\, A+H_{\mathrm{enviroment}}(p,\epsilon_{1},\cdots\,,\epsilon_{n})
\]
where $g(t)$ is a function (non vanishing during the measurement time), $A$ the
measurement operator and $p$ a pointer state which changes during the
measurement and anchors  down its properties in a ,,macroscopic``
number of tracers in the witnessing subsystem $\epsilon_{1},\cdots\,,\epsilon_{n}$. 

As in open systems the decoherence concept (einselection\,\cite{joos2013decoherence,zeh1993there})
plays a central role. It describes how a quite classical description
is reached by restricting the consideration to a local subsystem.
The trick is to dislocate unavoidable entanglements to unconsidered
remote parts. 

For a subsystem in a huge surrounding there are obviously a large
number of measurement processes as there is usually
no shielding from such interactions. Estimates showed\,\cite{joos2013decoherence,Kiefer2008siegen,zeh1993there}
that the bulk of the interactions does not change the state of the
considered system but just introduces a phase\,%
\footnote{Group theoretical both subsystems have the structure of a sub algebra
of the Lie algebras $SU(n_{\mathrm{measured\, subsystem}})$ and $SU(n_{\mathrm{wittnessing\, subsystem}})$.
The combined system $SU(n_{\mathrm{measured\, subsystem}}+n_{\mathrm{wittnessing\, subsystem}})$
contains among many other elements a $U(1)$ allowing for a arbitrary
relative phase between the subsystems which can be transferred to
the measured subsystem.%
} and a localization. The outcome is, that off-diagonal contributions
in local density matrices will effectively disappear. Coexisting essentially
coarse grained classical paths entangled to remote parts will remain.

Einselection does not help to select the actual classical path, but
the selections no longer have to originate in interactions within the
local system. Our concept is that in a second step the  multi-world
structure entangled to remote areas can be eliminated (collapsed)
by a projection to a given fixed final state.

In a two boundary system intermediate measurements have to be conditional
and yield the so called ,,weak value``\,\cite{aharonov2008quantum}:
\[
<A>_{\mathrm{weak}}=<\mathrm{initial\, state}\,|A|\,\mathrm{final\, state}>/<\mathrm{initial\, state\,}|\,\mathrm{final\, state}>
\] For non degenerate eigenvalues $a_{i}$ the probability that $A$ finds 
an intermediate state $a_{k}\:$is:
\[
P(a_{k})=|<\mathrm{initial\, state}\,|a_{k}>\,<a_{k}|\,\mathrm{final\, state}>|^{2}\,/\,|<\mathrm{initial\, state\,}|\,\mathrm{final\, state}>|^{2}
\]
The phase averaging eliminates interference contributions and the denominator
can be simplified to:
\begin{eqnarray*}
 & \sum_{i,j}<\mathrm{initial\, state\,}|a_{i}>\,<a_{i}|\,\mathrm{final\, state}><\mathrm{final\, state\,|a_{j}>\,<a_{j}|}\,\mathrm{initial\, state}>=\\
 & =\sum_{i,}|<\mathrm{initial\, state\,}|a_{i}><a_{i}|\,\mathrm{final\, state}>|^{2}
\end{eqnarray*}
yielding the Aharonov-Bergman-Lebowitz equation\,\cite{aharonov1964time}. 

Their picture is two symmetric evolutions one from the initial state
and one from the final one matching at the measurement times and fixing
the measurement outcome. The choice in the matching time 
\[
U(t_{\mathrm{initial}},t_{\mathrm{final}})=U(t_{\mathrm{initial}},t_{\mathrm{match}})U(t_{match},t_{\mathrm{final}})\,\,\,\,\,\,\,\mathrm{with}\,\, t_{match}\in[t_{\mathrm{initial}},t_{\mathrm{final}}]
\]is, of course a question of convenience. Except chapter 7 we use
$t_{match}=t_{\mathrm{final}}$ so that the so called quantum collapses are encountered
by projecting out the fixed final state. 

As the matching replaces a huge number of measurement collapses it will
naturally yield a tiny value (evaluated as ,,tiny`` in Zeh's book\,\cite{zeh2001physical}
for a big bang/crunch scenario). In contrast to Schulman's statistical
two boundary concept\,\cite{schulman1997Time} the fixed final state
doesn't select ,,special`` initial states. The density matrix of the
initial state is not taken as origin of the rich structure of
the universe.  Also the  wave functions are
physical objects with no hidden properties. 

Weak values can be unreasonable. The concept for their values to conform
with usual quantum predictions relies on the statistical assumption
that (in the Schr{\"o}dinger picture):

\begin{eqnarray*}
 & |<\mathrm{initial\, state\,|}U(t_{1}-t_{0})|\, a_{i}>/<\mathrm{initial\,
 state\,}|U(t_{1}-t_{0})|a_{j}>|=\\[3mm]
=\lim_{(t_{2}-t_{1})\to\infty}( & \,\,\,\,\,|<\mathrm{initial\, state\,}|\, U(t_{1}-t_{0})\,|\mathrm{\, projection}(a_{i})\,|\, U(t_{2}-t_{1})\,|\mathrm{\,\mathrm{final\, state}}>| &
 \: / \\
 & \,\,\,\,\,|<\mathrm{initial\, state\,}|\, U(t_{1}-t_{0})\,|\mathrm{\, projection}(a_{j})\,|\, U(t_{2}-t_{1})\,|\mathrm{\,\mathrm{final\, state}}>| & \,\,\,\,\,\,\,\,)
\end{eqnarray*}
The idea is that in a huge system and a long evolution time $t_{2}-t_{1}$
all intermediate states will find their matches with essentially equal
probability. 

Let us consider one example. The probability of an emission of a photon $e \to e + \gamma$
is according to the numerator of the first line proportional to  $\mathrm{e}^2$. If the photon is not 
contained in the final state it has to be captured again and its emission 
probability is therefore proportional to  $\mathrm{e}^4$\, 
\footnote{Such an emission and absorption were required in theories in which
the photon had no separate reality\,
\cite{tetrode1922wirkungszusammenhang,wheeler1945interaction,sussmann1952spontane}
A generalization is the "transactional"
interpretation~\cite{cramer1980generalized,cramer1986transactional}.
}. 
This doesn't contradict the above equation. 
The concept is that the 
situation is rich enough to offer many absorptive channels.
In this way the absorption probability $\sum_\mathrm{channels}
\mathrm{e}^2$ can be  unity\,
\footnote{The unity argument follows \cite{sussmann1952spontane}}.
There is no intrinsic distinction between emission and absorption in the argument.

 Aharanov et al.\,\cite{aharonov:2014zta}
showed that fixed initial and final states reclaims determinism. 
To avoid determinism slight variation in the the boundary states have to be admitted.

\section*{5\hspace*{0.5cm}Cosmology and the time arrow}
\vspace{-3mm}

As the closed system is not truly random it does not contain a time
arrow. Instead one can expect a complicated mesh of phases correlated
over large time distances. 

However as Gold observed\,\cite{gold1967nature} the special boundary
conditions of our expanding cosmos are quite consequential. It is 
counter intuitive as quantum mechanical processes and the cosmos
involve quite different scales. It is somewhat reminiscent to the 
resolution of Olber's paradox.

In macroscopia the cosmological expansion arrow of time is coupled to the 
thermodynamic one~\cite{cramer1986transactional} p.e.~by 
allowing the emission of thermal radiation 
in the dark sky of the expanding universe. The not returned radiation 
allows stars to loose energy. It facilitates the formation
of cold stars, etc. 

The time interval of our two boundary quantum system is taken to be large
enough to include this expansion. It  is easy to see how the 
expansion  transfers to a quantum mechanical time arrow. 
All quantum decision are encoded in the environment.
In an expanding universe the environmental witnesses predominantly
live in the much richer future. A tiny local changes at the 
time $t_{A}$ will therefore mainly affect the future environment
not involving    the past one. Inversely at a time $t_{A}$ the environment
$\epsilon(t_{A})$ will  mainly reflect the past. 

How do local measurements  inherit the cosmic time arrow?   
In a measuring process local phase correlations are destroyed 
as in an open system. The not so local witnessing system 
will typically spread out fast making a rejoining of different
outcomes on a short time scale unlikely. It will involve distinct
thermal radiation and eventually it will be in contact with cosmic thermal 
radiation processes. The central point is that the entangled thermal cosmic radiation
is sufficiently separate from its partners so that the disentanglement
is done by an projection on the far away final state and not by rejoining
interactions. 

Rejoining interactions could lead to a large number of macroscopically
coexisting classical paths causing a intermediate Everett system with a
growing number of multi-worlds. Our hypothesis is that - in spite of many 
somewhat accidental choices - essentially only one coarse grained 
,,dominant path`` survives on a macroscopic scale. 
The concept of ,,dominant path`` will be clarified below. 

The cost of fixed boundary states is an in the end 
deterministic macroscopic world. Determinism
is, of course, common to classical theories. A philosophical complication
arises as the system includes us. To be consistent with the felt reality
one would need to keep some not predetermined genuine ,,free
will``\,\cite{bohr1958physik,conway2006free,pittphilsci11893}.
Within the framework of physics an unpredictable piece of reality
seems prerequisite for such an influence. However this is not an argument
for canonical quantum mechanics with its unpredictable random collapses.
In this context there is no distinction between physically unpredictable
collapses  and unpredicted slight variations in the future boundary state. 
In both cases outside interventions can't be excluded as long as they stay  
on a statistically insignificant level.

\section*{6\hspace*{0.5cm}The time-ordered causal macroscopia}

How can one understand that such a theory with a fixed final quantum
state can coexist with a time-ordered seemingly causal interactive
world. To avoid the obvious contradiction one has to separate the
,,\textit{far future}`` macroscopia from the ,,\textit{near future}``
macroscopia we are usually concerned with. The conjecture is that
if the fixed final state is far away enough it cannot control our
,,\textit{near future}``. 

To illustrate the basic argument we consider a ,,router`` which
is a slightly random program which calculates the best way for a car
to come from position $A$ to position $B$~. In the ,,router``
program it should be possible to enforce an intermediate position
$A'$ and at such an intermediate positions the velocity and the direction
should be taken into account. 

We consider the case where the final position B is very far away from
a narrow region around $A$ and $A'$. The route behind $A'$ will
then within this region strongly depend on the position $A'$ and
on the velocity and direction at this point. The location of $B$
will be at first almost irrelevant. In this way the required fixed
outcome seems not to concern the nearby region. 

We now consider tiny variations of $A'$. Between the closely neighboring
$A$ and $A'$ there will be a correspondingly tiny change. As the
path from $A'$ and $B$ will have many options at furcation points
considerable changes can be caused by tiny variations in $A'$. In
this way the different scales of the distances introduce something
like a causal forward direction. 

To formulate the idea more precisely we used a toy program. In a two
dimensional space-time array we follow a moving object. 
Usually with a probability of $1-2\epsilon$
it moves:

\begin{center}
\begin{tabular}{|c|l|}
\hline 
two left & ~  if it moved two left one step before\tabularnewline
\hline 
one left & ~  if it moved one left one step before\tabularnewline
\hline 
 straight & ~ if it moved stright one step before ~ \tabularnewline
\hline 
~ one right ~ & ~ if it moved one right one step before\tabularnewline
\hline 
~ two right ~ & ~ if it moved two right one step before\tabularnewline
\hline 
\end{tabular}
\end{center}

\noindent 
and each with a probability $\epsilon$:

\begin{center}
\begin{tabular}{|c|}
\hline 
~as above, but  +1 resp -1~ \tabularnewline
\hline 
\end{tabular}
\end{center} 

\noindent The lattice is taken periodic to avoid edge effects. Steps are
limited to two units movements in $\delta x$. 
The algorithm is time symmetric. We fix the initial
position to $x=0$ and $\frac{dx}{d\delta t}|_{x=0}=1$ and using
rejection the final potion to $x=0$ and $\frac{dx}{d\delta t}|_{x=0}=0$.
With 1$0^{9}$ tries and 84663 (for the last time step weighted) counts 
we obtain the density
profile of the reached points shown in figure~2. For the random number
generator we use the method of \cite{marsaglia1990toward} implemented by
K. Hahn. A degree of causality
and retro causality for the near future or past is clearly seen. 
The direction of the initial
resp. final state is maintained for a few steps. The causal influence
vanishes for central times.

\begin{figure}[h]%
\includegraphics[scale=0.35]{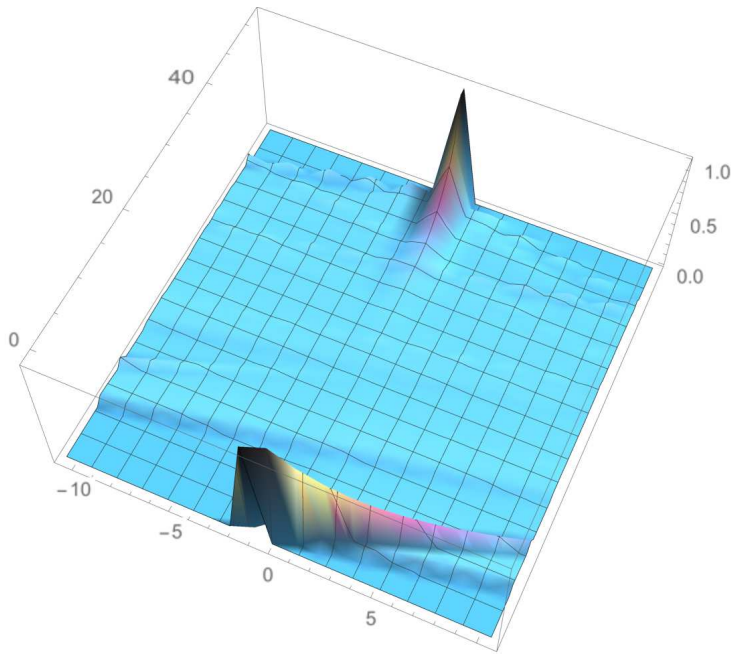}\protect\caption{Density
profile in a symmetric world}
\end{figure}%

Can such a situation result from quantum physics? 

The randomness in the toy model reflects the influence of unconsidered
radiation  constrained by the unknown final boundary state.
Einselection explains why classical paths (collections of quantum
paths) appear. To argue why typically one coarse grained path dominates 
we consider a bad double-slit experiment shown in figure~3. 

\begin{figure}
\noindent \begin{centering}
\includegraphics[scale=0.28]{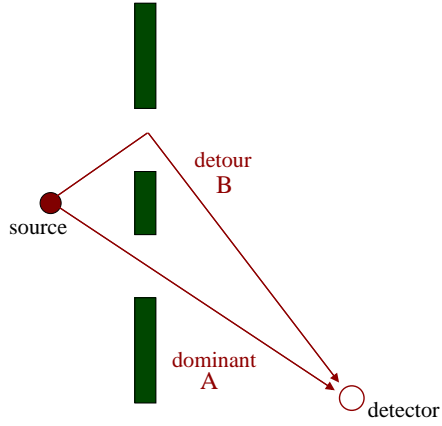}\protect\caption{%
A bad double-slit experiment%
}
\par\end{centering}
\end{figure}
The intensity of the contribution of a light source (indexed as ``o'') through
slits (indexed as ``$s$'' or separate as ``$s_{A}$'' and ``$s_{B}$'') in the detector
(indexed as ``d'')
in a two dimensional configuration with the plain of the slits at
$y_{s}=0$ is:
\[
I=|A+B|^{2}\propto |\int dx_{o}\, dx_{s}\, dx{}_{d}\,\exp(ik\cdot\mathrm{exponent})\:|^{2}
\]
The light with wave vector $k$ passing through lower slit contributes with a weight $A$
to the amplitude, the light passing through the upper slit with a
weight $B$. Defining $x_{s}=\tilde{x}_{s}+\Delta$ to first order
of $\Delta$ one obtains:
\[
\mathrm{exponent}=\mathrm{const}.+\left(\frac{x_{o}-\tilde{x}_{s}}{\sqrt{(x_{o}-\tilde{x}_{s})^{2}+y_{o}^{2}}}+\frac{\tilde{x}_{s}-x_{d}}{\sqrt{(\tilde{x}_{s}-x_{d})^{2}+y_{d}^{2}}}\right)\cdot\Delta
\]

For the slit $s(A)$ the coordinate $\tilde{x}_{s(A)}$ can be chosen
that the round bracket vanishes and in absence of an oscillating phase
the contribution $A$ will be proportional to the size of the slit
$s(A)$. 

Using distance $h$ between the slits we define $\tilde{x}_{s(B)}=\tilde{x}_{s(A)}+h$
and obtain for the slit $s(B)$:
\[
\mathrm{exponent}=\mathrm{const}.+\left(\frac{h}{\sqrt{(x_{o}-\tilde{x}_{s(B)})^{2}+y_{o}^{2}}}+\frac{h}{\sqrt{(\tilde{x}_{s}-x_{d(B)})^{2}+y_{d}^{2}}}\right)\cdot\Delta
\: .\]
As all distances are large against $1/k$ the integral $\int_{s(2)}d\Delta\,\exp(ik\cdot bracket\cdot\Delta)$
contains an oscillating phase massively reducing its contribution
to the amplitude (i.e. $|B|\ll|A|$) .

The decoherence effects and the extent of source e.c.t. eliminates interference
contributions between both classical paths. Two coarse grained ,,classical``
contribution a ,,\textit{dominant path}`` ($\propto|A|^{2}$) and
a tiny ,,\textit{detour-path}`` ($\propto|B|^{2}$) remain.

The appearance of a ,,\textit{dominant path}`` and a almost negligible
,,\textit{detour-path}`` seems typical. The conjecture is that during
the evolution such choices will repeat themselves over and over again.
Given the required final state an essentially single macroscopic ,,\textit{dominant
global path}`` is postulated.

In principle there are two kind of interactions. ,,Protected``\,\cite{aharonov2008quantum}
measurement processes leave the classical path unaffected, furcation
processes allow for alternate classical paths. As in the ,,router`` picture
tiny changes in the initial condition can change the way these
furcation points are met and so eventually lead to a completely different classical
path. This resurrects forward causality in the near future macroscopia.

The absence of backward causality might require that our expanding universe
is close to its initial state. It somehow fits with the presently
non decreasing expansion rate. We do not agree with Zeh (5.3.3)\,\cite{zeh2001physical}
that this means ,,improbably young``. 

To conclude this part we restate our conjecture:

$\bullet$~~~the influence of the not-open-ended nature is weak
enough not to disturb the seemingly causal near future of macroscopic
physics, 

$\bullet$~~~but strong enough to limit coexisting evolution paths
to a microscopic level typical for quantum phenomena. 

The initial and final states do not have to be pure states. However, the richness 
of the choices is taken to originate in the evolution. Even for pure states 
if the match 
\[
|<\mathrm{initial\, state}\,|S(t_i ,t_f)|\,\mathrm{final\, state}>|^{2}
\]
happens to be really tiny irregular disturbances will play an enhanced role. This
should increase the number of furcation points. In this way the system
can be externally adjusted to meet the required conditions.

The conjecture offers a way to reconcile quantum mechanics with the known
part of classical
physics in a completely consistent way without invoking epistemological
limitations of our description of quantum collapses. That such a theory
might be realized we consider an important  point strongly limiting
the motivation to search for modified quantum theories.

\section*{7\hspace*{0.5cm}Cosmology and the matching final state}

The considered evolution from a tiny initial state to a huge arbitrary fixed 
final one is a simple and straight forward cosmological possibility.  It is 
tempting to consider more complex scenario's.  A persuasive concept could 
relay on the central role of the forward backward CPT symmetry.  Often what
is allowed by symmetry is realized.  An eventually at least for a while   
contracting phase of the universe with an inverted time is therefore quite
plausible.  For simplicity we here discuss a big bang / crunch theory.

Quantum cosmology is clearly not a simple field. Prerequisite for the
consideration is a unitary evolution including of non trivial general
relativistic structures~\cite{Hooft:2016itl}.

The time direction originating in radiating off witnesses into an expanding
universe is  reversed during the contracting phase.  However, after a CPT
transformation \[
<a(t_1)|U|b(t_2)>\,\,\,\,\stackrel{\mathrm{CPT}}{\to}\,\,\,\,<a(T-t_1)|U|b(T-t_2)>^{*}
\] (where T  is the lifetime of the universe) probabilities are identical~%
\footnote{No statement is necessary about
the observed tiny CP or the related T violation.  If attributed to p.e.  a
CP asymmetric baryon imbalance of vacuum
\,\cite{zeh2001physical,bopp:2010um,bopp:2011zz} it has nothing to do with 
a real time arrow or the CPT    
structure.  The vacuum might evolve or switch on the considered
asymmetry.   }.
Except for how quantum measurements are recorded the time direction
is not relevant.

The physics in the in-between phase is unknown.  The lack of expansion might
not be important as the density could be very low and if the period is   
sufficiently short the sky might stay a dark sink for radiations in both 
directions.  It is cold enough so that emitted and absorbed radiation are
insignificant.

A  big bang / crunch scenario was considered by Gell-Mann and Hartle\,
\cite{gell1994time}.  They were particularly interested in a symmetric
situation with a single density matrix describing initial and final state.
If such a scenario can be excluded from observations is not clear\,
\cite{gell1994time,craig1996observation}.

In the here considered collapse-less quantum dynamic identical density
matrices at both boundaries mean identical evolutions  
\footnote{
The description with
formally independent evolutions is redundant.  It suffices to consider just
the wave function $\phi(t)$ and eliminate the complex conjugate using   
$\phi^*(t)=\phi(T-t)$ where $T$ is the lifetime of the universe and the
asterisk is the conjugate.

The probability that a state $\phi(t_1)$ goes to $\phi(t_2)$ requires a
congruence of a transition $<\phi(t_1)| U(t_1,t_2) |\phi(t_2)>$ in the
expanding universe and a transition $<\phi(T-t_2)| U(T-t_2,T-t_1)
|\phi(T-t_1)>$ in the contracting one. It provides a beautiful  explanation
of bilinear Born probabilities in quantum mechanics.
}.
At the boundary all evolved multi-versa will match.
In such a scenario the role of the fixed final states to eliminate  
or reduce multi-versa is therefore completely lost.

To keep our central argument it is necessary to have at least very different
final and initial density matrices.
To consider their evolution of $\phi(t)$ and $\phi^*(T-t)$ one can   
formally  split the two fixed boundary system into two parts of again two
boundary ones each with   
one fixed and one matching intermediate boundary.
Following the time arrow one  replaces the expanding and contracting   
evolution by two separate evolutions.

All quantum decisions have to be encoded in the final state. In a big bang /
big crunch scenario this might be problematic as both states are limited.
Possibly the richness of the intermediate (matching) state helps to
sufficiently reduce entanglement between the expanding and contracting
world so that not much is changed if one replaced the fixed final states by   
matching ones. To explain the mechanism we denote
entangled connections by underlining
\[
\underline{ab}=a'b'+a''b''
\]
and write an exemplaric  final state of the expanding universe as
\[
\underline{ab}cde\underline{fg}hijklm
\]
and that of the contracting one as
\[
ab\underline{cd}efg\underline{hi}jklm ~.
\]
As matching entanglements vanish in the limit of a huge intermediate state
all entanglement will get typically lost. 
The matching albeit not fixed final state
might suffice to avoid multi-verses.

\section*{Conclusion}

To conclude heuristic arguments are given that dropping time ordered
causality for quantum states needs not destroy the apparent causality
on a macroscopic level. If the conjecture is correct one would obtain
a consistent synthesis of microscopic and macroscopic physics. A more
formal, less heuristic description would clearly be helpful.

\section*{Acknowledgment}

We thank David Craig, Claus Kiefer and Wolfgang Schleich for help in pointing out
relevant literature. 


\bibliography{Ncausal_QM_52}

\end{document}